\documentclass{article}
\usepackage{spconf,amsmath,graphicx,amsfonts}
\usepackage{algorithm}
\usepackage{algorithmic}
\usepackage{multirow}
\usepackage{amsmath} 
\usepackage[misc]{ifsym} 

\usepackage{enumitem}
\setlist{nosep, leftmargin=14pt}

\usepackage{mwe} 

\title{STPDnet: Spatial-temporal convolutional primal dual network for dynamic PET image reconstruction}
\name{Rui Hu\textsuperscript{1}, Jianan Cui\textsuperscript{1}, Chengjin Yu\textsuperscript{1}, Yumei Chen\textsuperscript{2}, Huafeng Liu\textsuperscript{1}$^{(\textrm{\Letter})}$}
\address{\textsuperscript{1}State Key Laboratory of Modern Optical Instrumentation, Department of Optical Engineering,\\Zhejiang University, Hangzhou 310027, China\and \textsuperscript{2}Department of Mathematics, University of Florida, Gainesville, FL 32611 USA}

\begin{document}

\maketitle
\begin{abstract}
Dynamic positron emission tomography (dPET) image reconstruction is extremely challenging due to the limited counts received in individual frame. In this paper, we propose a spatial-temporal convolutional primal dual network (STPDnet) for dynamic PET image reconstruction. Both spatial and temporal correlations are encoded by 3D convolution operators. The physical projection of PET is embedded in the iterative learning process of the network, which provides the physical constraints and enhances interpretability. The experiments of real rat scan data have shown that the proposed method can achieve substantial noise reduction in both temporal and spatial domains and outperform the maximum likelihood expectation maximization (MLEM), spatial-temporal kernel method (KEM-ST), DeepPET and Learned Primal Dual (LPD).
\end{abstract}
\section{Introduction}
\label{sec:intro}
Dynamic Positron Emission Tomography (dPET) reconstruction can provide physiological and metabolic information through the \textit{in vivo} spatial-temporal distribution of the radiolabeled tracer, which can be used to detect and characterize a variety of diseases such as heart diseases and cancer\cite{gambhir2002molecular}. However, the count of dynamic data in a single frame is lower than that of static data, especially in the early period of the scan, resulting in severe noise in the reconstructed images, which limits the clinical application of dPET to a large extent.

For the reconstruction of the individual time frame, traditional algorithms include analytic methods, iterative methods, and penalized likelihood methods. The analytic method such as the filtered back projection (FBP\cite{brooks1976statistical}) suffers from substantial noise and strip artifacts. Iterative method such as maximum likelihood expectation maximization (MLEM\cite{shepp1982maximum}) improves the image quality compared with FBP, but the reconstruction result is limited by the accuracy of the projection model. Penalized likelihood methods\cite{chen2015sparse,xie2020penalized} further optimize the reconstruction through regularization. Nevertheless, the choice of regularization is an open question. To explore both spatial and temporal correlations, spatial-temporal kernel method (KEM-ST\cite{wang2018high}) achieves substantial noise reduction. However, the global temporal kernels may not match the local temporal characteristics in some atypical regions and the large temporal window size may result in oversmoothed images. 

Deep learning methods are widely applied to PET image reconstruction and show better reconstruction results than traditional methods\cite{gong2019machine,cui2019structure}. Most deep learning techniques in PET are used for post-processing\cite{cui2019pet}. However, post-processing can not save the lengthy reconstruction time and its results are sensitive to the pre-reconstruction algorithm. Direct learning the sinogram to image mapping is a popular way for PET image reconstruction\cite{haggstrom2019deeppet,li2022deep}, but lacking the constraints of the physical projection matrix makes this strategy data-hungry and having poor generalization. Model based deep learning methods\cite{mehranian2020model,hu2022transem} show inspiring results in both generalization and interpretability, which has been a promising solution.

Whereas, the deep learning based methods mentioned above rarely consider the temporal relation between the different frames of the measured sinogram data, which potentially affects the accuracy of the reconstructed time activity curves (TACs) and the resolution of the reconstructed images.

In this paper, we propose a spatial-temporal convolutional primal dual network (STPDnet) for dynamic PET image reconstruction. Spatial-temporal 3D convolution enables our method simultaneously model the spatial and temporal correlations in dynamic PET measured sinograms. Combined with physical projection constraints, STPDnet has good interpretability and stability. Besides, STPDnet achieves substantial noise reduction in both spatial and temporal domains in the experiments of the rat data sets.

\section{Methods}
\label{sec:format}

\begin{figure*}                
\centering
\includegraphics[width=\textwidth]{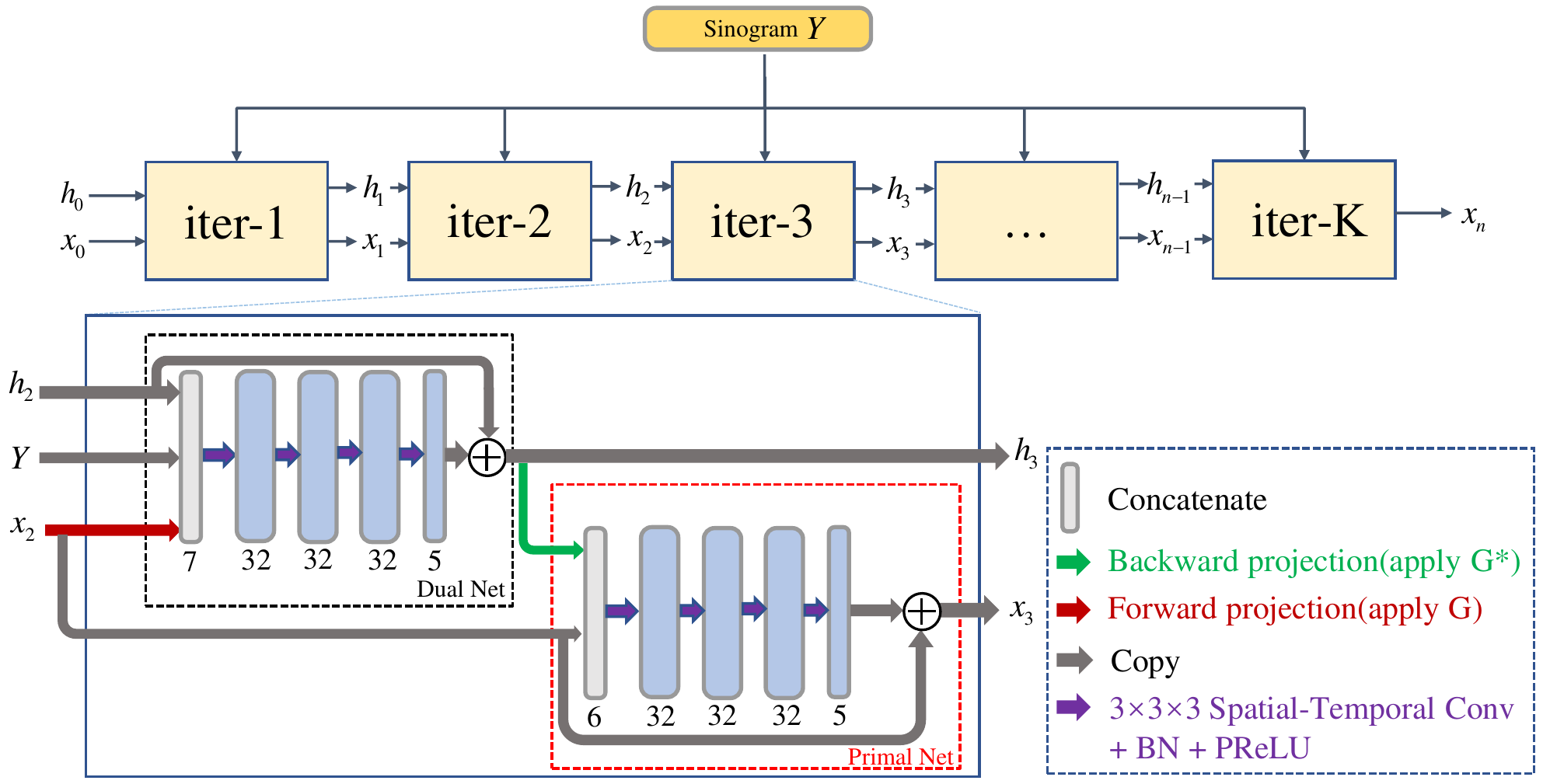}  
\caption{The network structure of the proposed spatial-temporal primal dual network (STPDnet).The whole reconstruction process is unrolled into several iterative modules, each module contained a dual net (D) for the updating of the dual variable $h$ and a primal net (P) for the updating of the primal variable $x$.}
\label{Fig1}                     
\end{figure*}

\subsection{Dynamic PET image reconstruction}
In dynamic PET image reconstruction, for each time frame $t$ , the expectation of the projection data ${\bf{\bar y}} = \{ {\bar y_{i,t}}\}$ can be modeled by 
\begin{equation}
\label{eq1}
    {\bf{\bar y}} = G \cdot {\bf{x}} + {\bf{r}}
\end{equation}
where ${\bf{x}} = \{ {x_{j,{t}}}\}$ is the unknown image, $G\in {{\mathbb{R}}^{I\times J}}$ is the system response matrix, ${G_{i,j}}$ represents the probability that a photon emission form $j$-th voxel received by the $i$-th detector, which depends on the physical properties of PET scanners. $\bf{r}$ is the expectation of dynamic scattered and random events.

Minimizing a regularized objective function is a common model driven approach for solving Eq.\ref{eq1}:
\begin{equation}
\label{eq2}
\mathop {\min }\limits_{{\bf{x}} \in X} L({\bf{G \cdot x}},{\bf{y}}) + \lambda R({\bf{x}})
\end{equation}
where $L({\bf{G \cdot x}},{\bf{y}})$ is the data fidelity term that usually computed using the negative log-likelihood and $R({\bf{x}})$ is the regularization term. $\lambda$ is the regularization parameter. 

\subsection{Spatial-temporal convolutional Primal Dual network}
\subsubsection{Learned primal dual}
The optimization problem like Eq.\ref{eq2} can be solved by the Learned Primal Dual(LPD)\cite{adler2018learned} network in two iterative steps with a dual variable $\bf{h}$:
\begin{equation}
\label{eq3}
    {{\bf{h}}_{k}} \leftarrow {\Gamma _{\theta _k^d}}({{\bf{h}}_{{k - 1}}},{\bf{G}} \cdot {{\bf{x}}_{{k - 1}}},{\bf{y}})
\end{equation}
\begin{equation}
\label{eq4}
    {{\bf{x}}_{k}} \leftarrow {\Lambda _{\theta _k^p}}({{\bf{x}}_{{k - 1}}},{{\bf{G}}^{\bf{*}}} \cdot {{\bf{h}}_{k}})
\end{equation}
where $\bf{h}\in{Y}$ (measurement sinogram space), $\bf{x}\in{X}$ (reconstructed image space) and $k$ denotes $k$-th iteration. ${\bf{G}^*}$ is the adjoint of the operator $\bf{G}$. ${\Gamma _{\theta _k^d}}$ and ${\Lambda _{\theta _k^p}}$ are two learned proximal ($d$ for dual, $p$ for primal) with the parameter ${\theta _k^d}$ and ${\theta _k^p}$. 

\begin{figure*}                
\centering
\includegraphics[height=8.2cm,width=\textwidth]{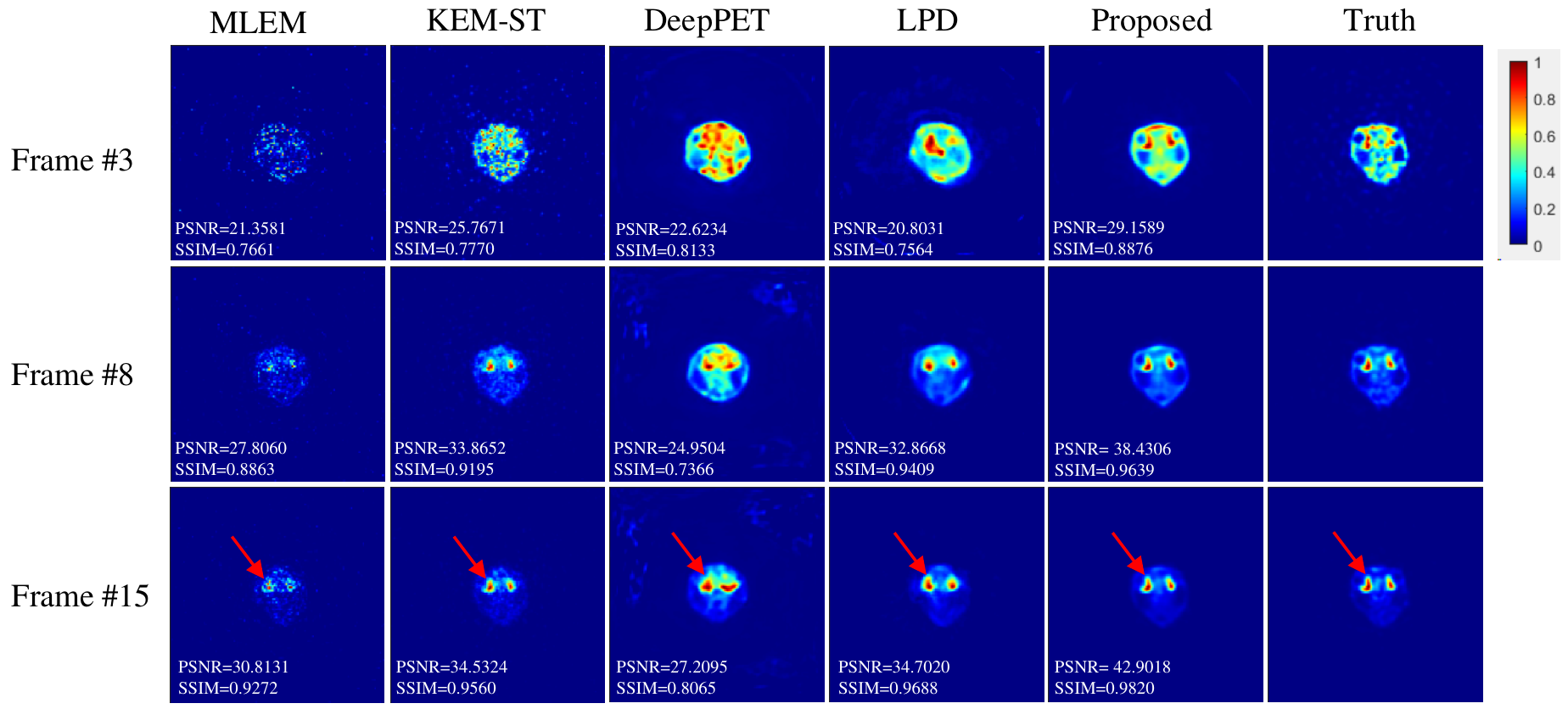}  
\caption{Reconstructed images by different reconstruction methods for the 3th frame (top row), 8th frame (middle row) and 15th frame (bottom row). From left to right: MLEM, Spatial-temporal kernel method, DeepPET, Learned primal dual, proposed STPDnet and the ground truth.}
\label{Fig2}                     
\end{figure*}

LPD has been successfully used in low-dose CT(LD-CT)\cite{adler2018learned} and Compressed Sensing MRI(CS-MRI). For CT and MRI, the modeling of the operator $G$ is accurate and common. Moreover, the spatial resolution is the only consideration in most cases. While in dynamic PET reconstruction, noise suppression in the spatial domain is not the only thing we need to focus on. The temporal dependency of the reconstructed images especially the accuracy of the time activity curve(TAC) is equally important for kinetic modeling and analysis. In addition, unlike CT and MRI, it is often difficult to have an accurate estimate of the system response matrix $G$ in PET, which potentially affect the quality of the reconstructed images.
\subsubsection{Proposed method}
To capture both the temporal and spatial information of dynamic PET measurement data, we propose a spatial-temporal convolutional primal dual network (STPDnet) as shown in Fig.\ref{Fig1}. The whole reconstruction process is unrolled into $n$ iterative modules, each module contained a dual net ($D$) for the updating of the dual variable $\bf{h}_{i,t}$ and a primal net ($P$) for the updating of the primal variable $\bf{x}_{j,t}$. The 3D spatial temporal convolution with kernel size 3$\times$3$\times$3 is adopted in each primal net and dual net, the time correlation between the primal net and dual net does not interacting, and the batch normalization (BN) is introduced behind every convolution operator to enable the network to learn more efficiently. In order to reduce the effect of low accuracy of projection operator on reconstruction results, we add a convolution layer after each projection to learn the gap between the simulated PET projection and the real world projection. Techniques like BN and skip connection allow the network to be much deeper with better generalization ability and make it particularly suitable for the dynamic PET image reconstruction. The overall algorithm flowchart is presented in Algorithm 1.
\begin{algorithm}
\caption{Algorithm for dynamic PET reconstruction with spatial-temporal convolutional Primal Dual net-work} 
\label{alg1} 
\renewcommand{\algorithmicrequire}{\textbf{Input:}}
    \begin{algorithmic}[1] 
    \REQUIRE ~~Image initialization ${\bf{x}}_{j,t}^0$, dual variable initialization ${\bf{h}}_{i,t}^0$, maximum number of iteration blocks $K$, measured Sinogram $\bf{y}$
    \FOR{$k \in [1,K]$}
    \STATE ${{\bf{h}}_{i,t}^{k}} \leftarrow {D}({{\bf{h}}_{i,t}^{k-1}},{\bf{G}} \cdot {{\bf{x}}_{j,t}^{k - 1}},{\bf{y}})$
    \STATE ${{\bf{x}}_{j,t}^{k}} \leftarrow {P}({{\bf{x}}_{j,t}^{k-1}},{{\bf{G}}^{\bf{*}}} \cdot {{\bf{h}}_{i,t}^{k}})$
    \ENDFOR
    \RETURN $\bf{x}^K$; 
    \end{algorithmic}
\end{algorithm}

\subsection{Implementation details and reference methods}
The STPDnet was implemented using Pytorch 1.7 on a NVIDIA TITAN-X. The number of iteration blocks is 10. The number of primal and dual variables is 3 determined by the ablation experiment. The image ${\bf{x}}_{j,t}^0$ and dual variable ${\bf{h}}_{i,t}^0$ were both initialized with values of zero. During training, the Adam optimizer was used and the mean square error(MSE) loss was calculated between the network outputs and the label images. The learning rate is 8e-4 and decays by a factor of 0.99 after each training epoch. The STPDnet was trained 200 epochs and batchsize was 2.
We compared our method with Maximum likelihood expectation maximization(MLEM)\cite{shepp1982maximum}, Spatial-temporal kernel method(KEM-ST)\cite{wang2018high}, DeepPET\cite{haggstrom2019deeppet} and Leaned primal dual(LPD)\cite{adler2018learned}. For both MLEM and KEM-ST, 20 iterations were used and the size of temporal windows in KEM-ST was 15. The learning rate was 5e-4, batchsize was 72 and epochs were 200 in the training of DeepPET. The parameter settings for training of LPD was the same as STPDnet.
\section{Experimental Results}
\label{sec:Results}
\subsection{Rat data acquisition}
Thirteen rats with gliomas data sets of one hour FDG dynamic scan acquired on Siemens Micro-PET/CT Inveon scanner with 1 mCi dose injection were employed in this study. Data acquisition began right after the FDG injection. The scanning schedule consisted of 18 time frames over 60 minutes: 3 $\times$ 60 $s$, 9 $\times$ 180 $s$, 6 $\times$ 300 $s$. Only the segments 0 of the michelograms (1/25 of the full counts) were used as training and test sinograms (160 views, 128 bins) for low count simulation.
For a single slice sinogram, frame 1 has 5k events, whereas frame 18 has 20k events approximately. These low-count sinograms were taken as inputs, the reconstructed PET images with CT attenuation correction and full 3D counts were used as labels. Ten rats were selected randomly for training, 1 for validation and 2 for testing. To ensure that the normalization does not change the temporal dependence of the input data, we divided the sinograms and labels by their maximum value of all frames.

\begin{figure}
\begin{minipage}{.48\linewidth}
  \centering
  \centerline{\includegraphics[width=4.0cm]{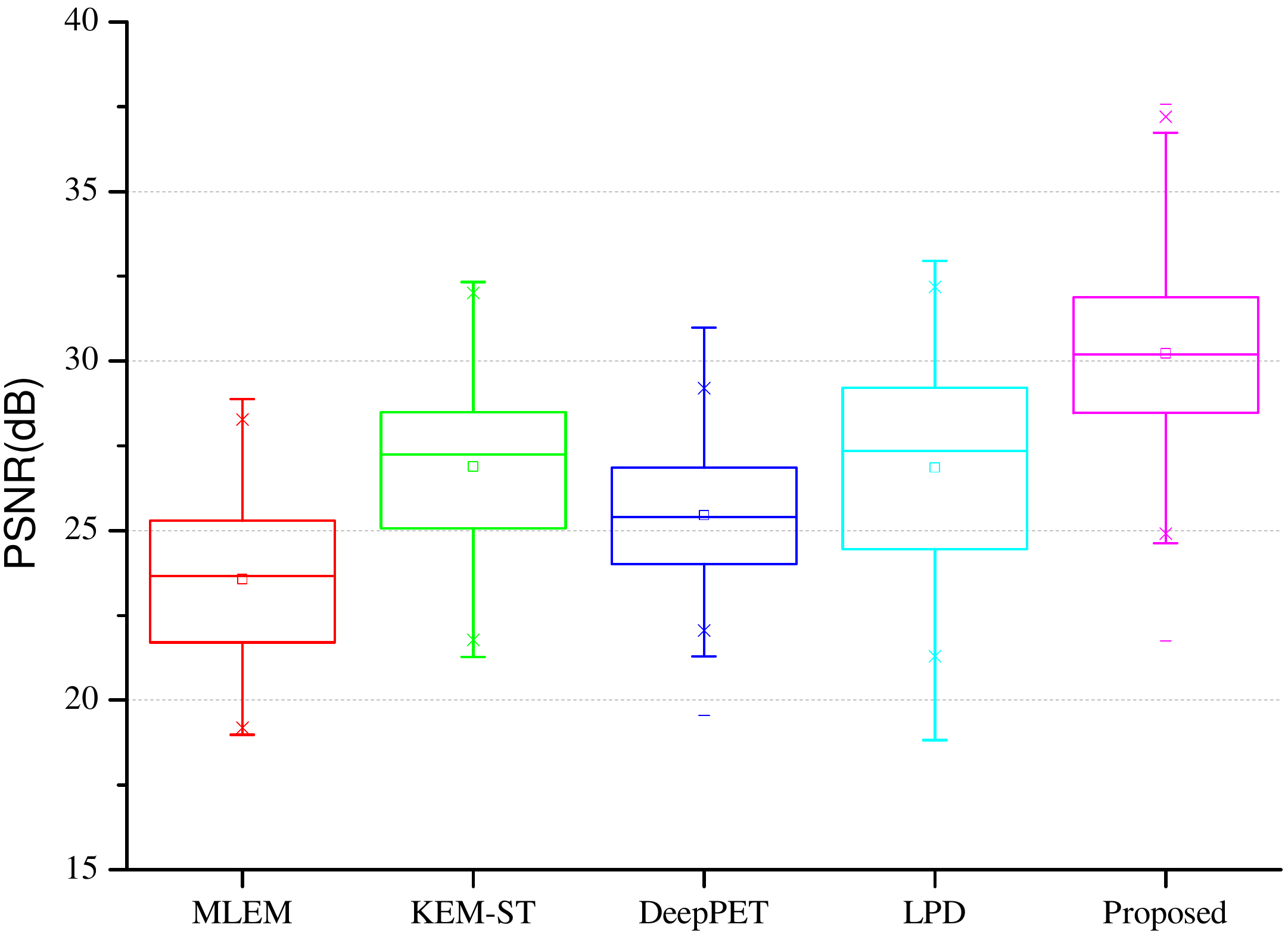}}
  \centerline{(a) PSNR}\medskip
\end{minipage}
\hfill
\begin{minipage}{0.48\linewidth}
  \centering
  \centerline{\includegraphics[width=4.0cm]{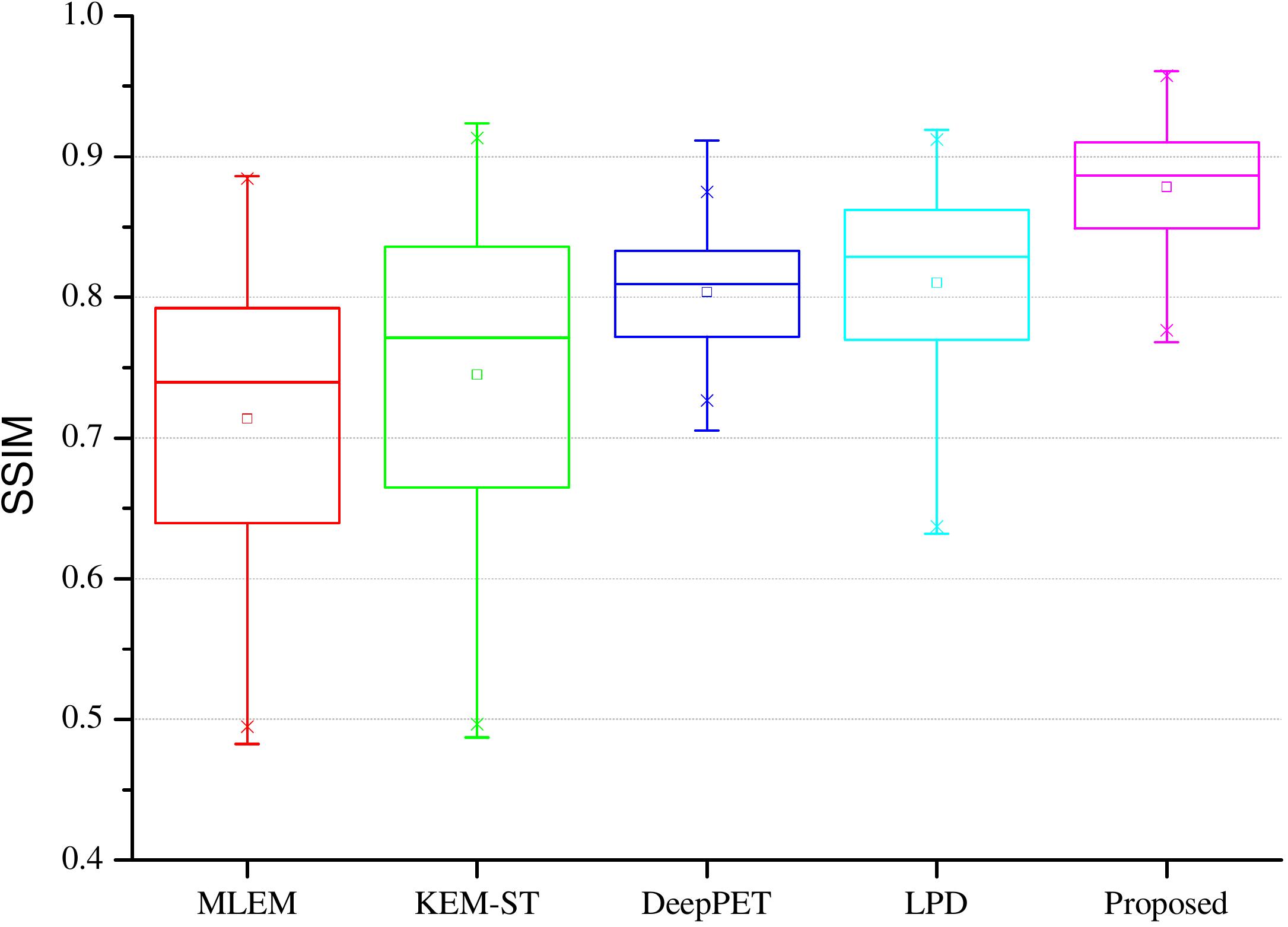}}
  \centerline{(b) SSIM}\medskip
\end{minipage}
\caption{Box plots of the mean PSNR and SSIM of all time frames of the test set reconstructed by different reconstruction methods.}
\label{Fig3}
\end{figure}

\subsection{Comparison for overall reconstructed image quality}
Fig.\ref{Fig2} shows the comparison of reconstruction images by the five methods for the 3th frame, 8th frame and 15th frame. In the case of low counts, without any additional prior information, MLEM obtained the lowest PSNR and SSIM and poor image quality. By incorporating temporal correlations, the KEM-ST improved the image quality, but there was still significant noise. As a totally data driven method, DeepPET showed an unstable results. The poor generalization resulted in the performance of DeepPET being worse than the traditional methods in some cases. Combined with the physical projection matrix constraints of PET and the learning ability of the network, LPD had a good improvement in PSNR and SSIM. However, due to lack of temporal information modeling, LPD performed poorly in tumor details. The proposed STPDnet beat all the comparison methods in both structures and details.
Fig.\ref{Fig3} shows the box plots of mean PSNR and SSIM of all frames of the test set. The PSNR and SSIM of images reconstructed by STPDnet were significantly higher than other methods over all time frames.
\subsection{Comparison for tumor time activity curves}

\begin{figure}[t]
\begin{minipage}{.48\linewidth}
  \centering
  \centerline{\includegraphics[width=4.0cm]{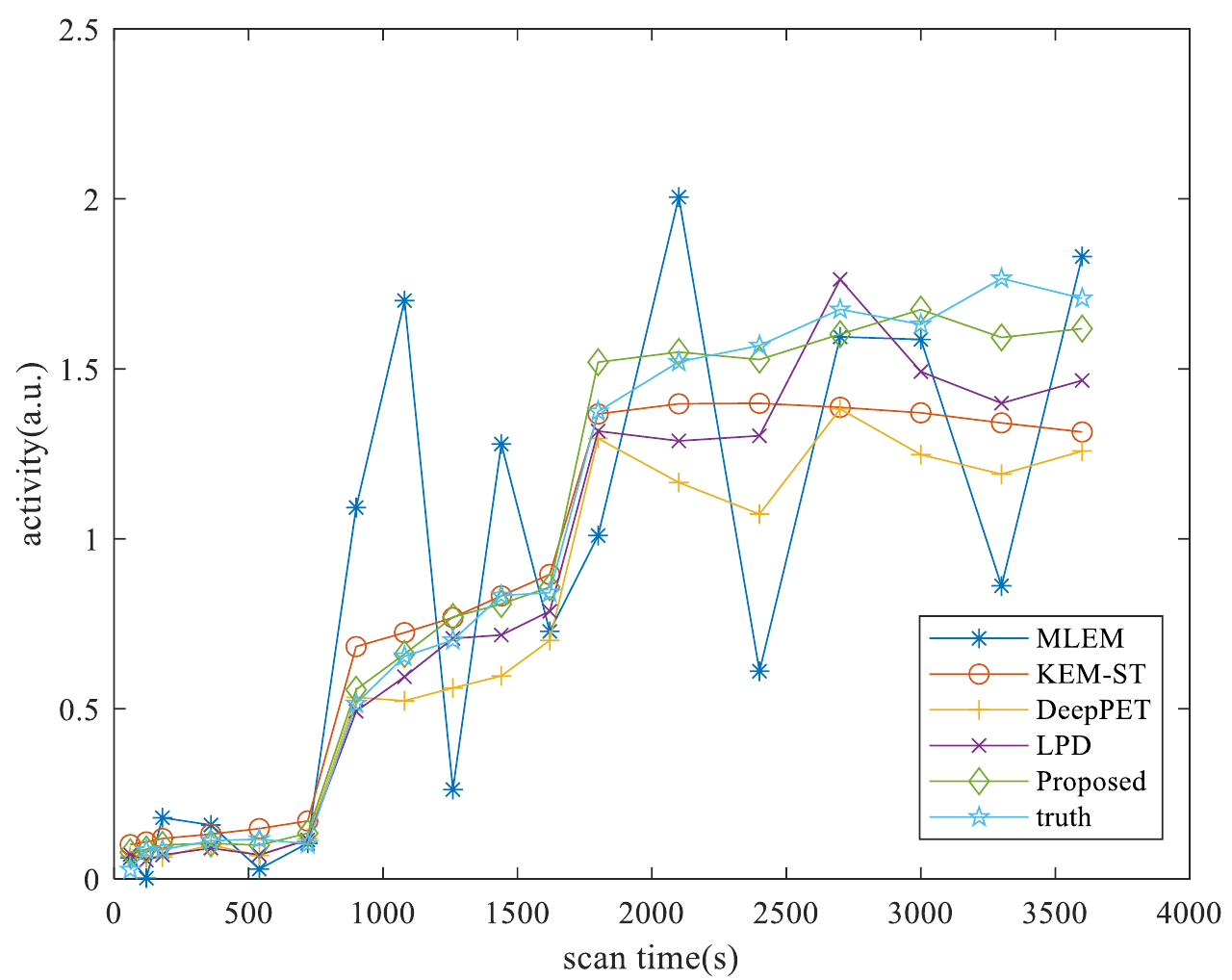}}
  \centerline{(a)}\medskip
\end{minipage}
\hfill
\begin{minipage}{0.48\linewidth}
  \centering
  \centerline{\includegraphics[width=4.0cm]{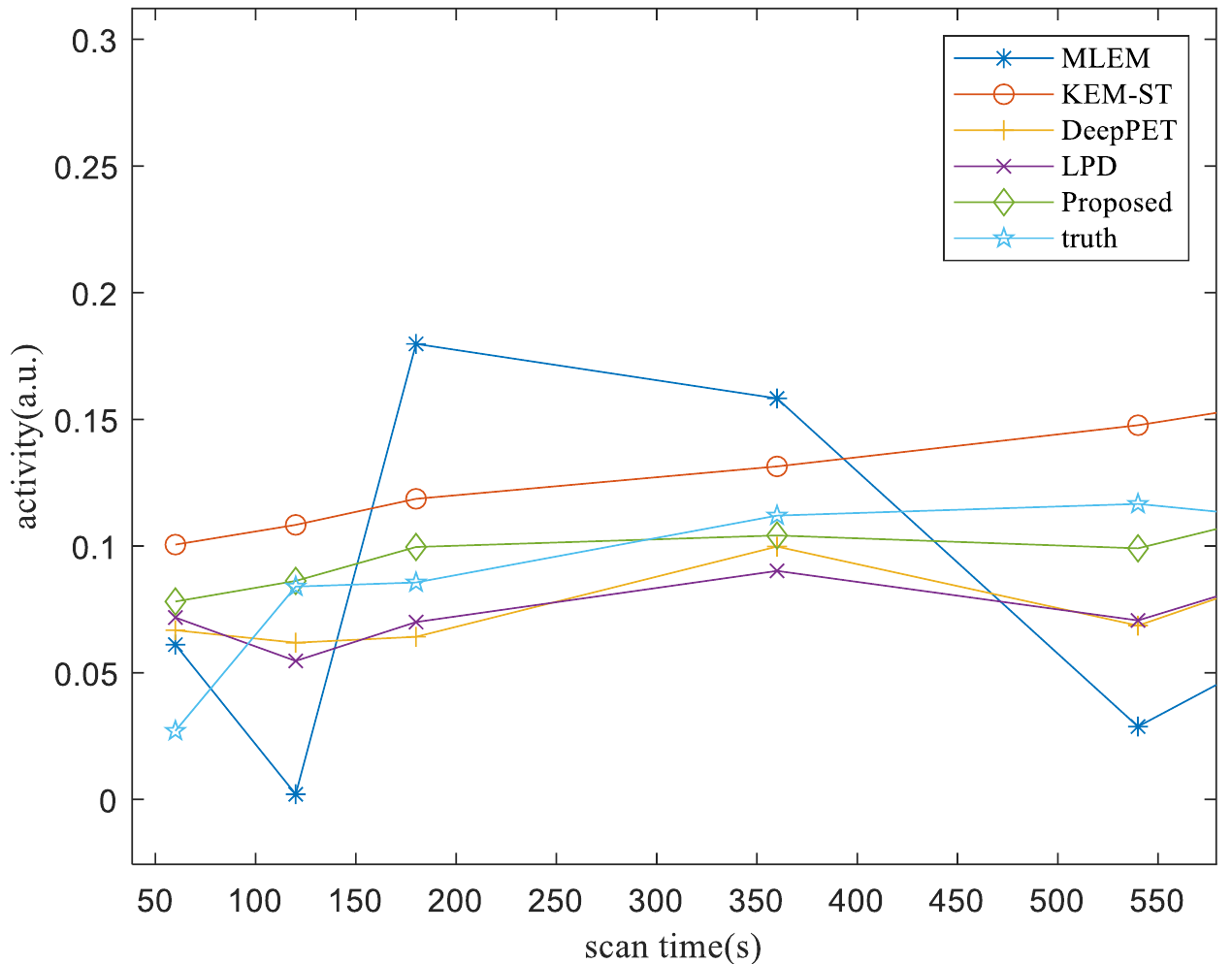}}
  \centerline{(b)}\medskip
\end{minipage}
\caption{Time activity curves of a tumor region reconstructed by different reconstruction methods. (a) tumor TAC, 0-60 minutes (1-18 frames), (b) tumor TAC, 0-9 minutes (1-5 frames).}
\label{Fig4}
\end{figure}

Fig.\ref{Fig4} shows the Time Activity Curves (TACs) of a pixel in the tumor region reconstructed by different reconstruction methods. The TACs of the first 5 frames were shown in Fig.\ref{Fig4}(b). The TAC reconstructed by MLEM fluctuated widely both in the first and last few frames. KEM-ST reduced temporal noise but over-smoothed in the late-time frames. The reconstructed TAC of DeepPET and LPD showed a certain degree of temporal noise because the temporal dependence of the data was not considered. In comparison, the STPDnet achieved a significant noise reduction in the temporal domain for both last-time frames and early-time frames.  

\section{Conclusion}
\label{sec:Conclusion}
In this paper, we have proposed a spatial-temporal primal dual network to learn both the spatial and temporal information of the measured data for low-count dynamic PET image reconstruction. The results from rat experiments have shown that the proposed method improved both spatial and temporal resolution and outperformed the mainstream methods. Future work will include more patient study.

\section{COMPLIANCE WITH ETHICAL STANDARDS}

This study was performed in line with the principles of the Declaration of Helsinki. Approval was granted by the Ethics Committee of Zhejiang university.

\section{ACKNOWLEDGMENTS}

This work is supported in part by the National Natural Science Foundation of China (No: U1809204,  61525106, 62101488), by the Key Research and Development Program of Zhejiang Province (No: 2021C03029) and by the Talent Program of Zhejiang Province (No: 2021R51004).

\bibliographystyle{IEEEbib}

\bibliography{ISBI2023.bib}

\end{document}